\documentstyle[prd,aps,twocolumn,epsfig]{revtex}
\def\br{\begin{eqnarray}}
\def\er{\end{eqnarray}}
\def\be{\begin{equation}}
\def\ee{\end{equation}}
\def\a{\alpha}

\def\d{\delta}
\def\g{\gamma}
\def\G{\Gamma}
\def\l{\lambda}
\def\L{\Lambda}
\def\m{\mu}

\def\l{\label}
\def\({\left(}
\def\){\right)}
\def\<{\left\langle}
\def\>{\right\rangle}

\def\S{\Sigma}

\begin{document}
\twocolumn[\hsize\textwidth\columnwidth\hsize\csname 
@twocolumnfalse\endcsname                            
%
%
\title{ Energy criterion to select the behavior of dynamical masses
in technicolor models }
\author{
A. Doff and A. A. Natale\\
}
\address{
Instituto de F\'{\i}sica Te\'orica, UNESP,
Rua Pamplona 145,
01405-900, S\~ao Paulo, SP,
Brazil}
\date{\today}
\maketitle
\begin{abstract}
We propose a quite general ansatz for the dynamical mass in technicolor models.
We impose on this ansatz the condition for formation of the tightest composite
boson state, or the criterion that it should lead to the deepest minimum of
energy. This criterion indicates a particular form of the technifermion self energy.
\end{abstract}


\vskip 0.5cm]

In the standard model of elementary particles the fermion and gauge boson
masses are generated due to the interaction of these particles with elementary
Higgs scalar bosons. Despite the success there are some points in the model
as, for instance, the enormous range of masses between the lightest and heaviest
fermions and other peculiarities that could be better explained with
the introduction of new fields and symmetries. One of the possibilities in this
direction is the substitution of elementary Higgs bosons by composite ones in the
scheme named technicolor\cite{kl}.

The beautiful characteristics of technicolor (TC) as well as its problems
were clearly listed recently by Lane\cite{lane}. Most of the technicolor
problems may be related to the dynamics of the theory as described in
Ref.\cite{lane}. Although technicolor is a non-Abelian gauge theory it is
not necessarily similar to QCD, and if we cannot even say that QCD is fully understood
up to now, it is perfectly reasonable to realize the enormous work that is needed
to abstract from the fermionic spectrum the underlying technicolor dynamics.

The many attempts to build a realistic model of dynamically generated
fermion masses are reviewd in Ref.\cite{kl,lane}. Most of the work in this area
try to find the TC dynamics dealing with the particle content of the
theory in order to obtain a technifermion self-energy that does not
lead to phenomenological problems as in the scheme known as walking
technicolor\cite{walk}. The idea of this scheme is quite simple. First,
remember that the expression for the TC self-energy is
proportional to   $ \Sigma (p^2)_{TC} \propto { (\langle \bar{\psi} \psi\rangle_{TC}}/{p^2})
(p^2/\L^2_{TC})^{-\gamma^*}$, where $\langle \bar{\psi} \psi\rangle_{TC}$ is
the TC condensate and $\gamma^*$ its anomalous dimension. Secondly, depending
on the behavior of the anomalous dimension we obtain different
behaviors for $ \Sigma (p^2)_{TC}$. A large anomalous dimension may solve the
problems in TC models. In principle we could deal with
many different models, varying fermion representations and particle content,
finding different expressions for  $ \Sigma (p^2)_{TC}$ and testing them
phenomenologically, i.e. obtaining the fermion mass spectra without
any conflict with experiment.

In this work we will introduce as one ansatz a quite general
expression for the technifermion self-energy. We will discuss its general properties
without paying attention to any group structure and will verify when this
ansatz imply that we have the tightest composite scalar boson, or the
deepest minimum of energy of the theory. In principle if we are able to
find the most probable expression for the technifermion self-energy based
on a general criterion we will only need to find the right theory that
lead to this formal expression.

In order to stablish a quite general ansatz for $ \Sigma (p^2)_{TC}$ we
go back to early work on the phenomenology of chiral symmetry breaking
in QCD, not because we shall assume that TC is similar to QCD, but because
the knowledge about solutions of the Schwinger-Dyson equations for the fermion
propagator in the QCD case will help us to find our general ansatz. This equation
as a function of the fermion and gauge boson propagators ($S$ and $D$ respectively)
is given by
\begin{equation}
S^{-1}(p)=\not\!{p}
-\imath \frac{4}{3} \int_{}^{\L} \frac{d^4q}{(2 \pi )^4}
\gamma_\mu S(q)\Gamma_\nu(p,q)g^2D^{\mu\nu}(p-q),
\label{e13}
\end{equation}
and has two asymptotic solutions\cite{lane2}
\be
\S_I(-p^{2}) = \mu[1 + bg^2(\mu^2)ln(-p^2/\mu^2)]^{-\g}
\label{si}
\ee
\be
\hspace{0.5cm}\S_R (-p^{2}) = \frac{\mu^3}{-p^2}[1 + bg^2(\mu^2)ln(-p^2/\mu^2)]^{\g} ,
\label{sr}
\ee
which are named respectively as irregular and regular solutions, where $\mu$
is the dynamical fermion mass ($\approx \langle \bar{\psi} \psi\rangle^{1/3} $),
and $\g = 3c/16\pi^{2}b$, with c given by
$$
c = \frac{1}{2}[C_{2}(R_{1}) + C_{2}(R_{2})- C_{2}(R_{3})]
$$
\noindent where $C_{2}(R_{i})$ is the Casimir operator for the fermions in the
representations $R_1$ and $R_2$ that condensate in the representation $R_3$,
and $b$ is the coefficient of $g^{3}$ in the $\beta$ function.
Only the solution $\Sigma_R (p)$ is compatible with OPE\cite{lane2} and is consistent
with asymptotic freedom. The $\Sigma_I$ solution can only be understood
if the theory has an explicit breaking of the chiral symmetry.

These solutions show naturally the extreme forms that we are looking for. One
obeys asymptotic freedom ($\Sigma_R (p)$), appears in the case of a perturbative anomalous
dimension, and lead to the known TC problems. Any other form of self-energy
decaying faster than $1/p^2$ is not interesting phenomenologically (i.e. it is worse
than $\Sigma_R (p)$). The other solution ($\Sigma_I (p)$) could only be understood if suitable
new interactions are assumed to be relevant at the scale of the cutoff of Eq.(\ref{e13}),
or if, alternatively, the ultraviolet cutoff is eliminated altogether as assumed in
the model of Ref.\cite{soni}. The only restriction on this solution is $\g > 1/2$\cite{lane2},
and if we consider the formal equivalence between the solution of the Schwinger-Dyson equation
with the Bethe-Salpeter one for pseudoscalar bound states, the above restriction indicates
the condition for wave function normalization of the Goldstone bosons.

Considering the above discussion we can formulate the following ansatz for $ \Sigma (p^2)_{TC}$
\be
\S(-p^{2})_{TC} = \mu\(\frac{\mu^2}{-p^2}\)^{\a}[1 + bg^2(\mu^2)ln(-p^2/\mu^2)]^{-\beta{cos(\a\pi)}}.
\label{sa}
\ee
This choice interpolates between $\Sigma_R (p)$ and $\Sigma_I (p)$. When $\a \rightarrow 1$
we reproduce Eq.(\ref{sr}) with $\beta = \g$, and when $\a \rightarrow 0$ the solution of
explicit chiral breaking is obtained. As far as we know there is not any solution that has
been discussed in the TC literature that cannot be represented by Eq.(\ref{sa}).

We can now discuss which are the basic conditions that Eq.(\ref{sa}) should satisfy.
Since $\S(p)_{TC}$ is also equivalent to the solution of the Bethe-Salpeter equation
for the scalar sector of technibosons, we could impose that the theory should be
stable when it forms the tightest bound states. This condition is the same as saying that
$\S(-p^{2})_{TC}$ must minimize the vacuum energy (or the vacuum expectation value of the
TC condensate).

To compute the vacuum energy for the technifermion self-energy we
can make use of the effective potential for composite operators
which is given by\cite{cornja}
\br
V(S,D) &=& - \imath \int \frac{d^4p}{(2\pi)^4}
Tr ( \ln S_0^{-1}S - S_0^{-1}S + 1) \nonumber \\
&& +\,\,V_2(S,D),
\l{vfull}
\er
where $S$ and $D$ are the complete propagators of
fermions and gauge bosons, respectively; $S_0$ and $D_0$, the corresponding bare
propagators.
\begin{figure}[ht]
\begin{center}
\includegraphics{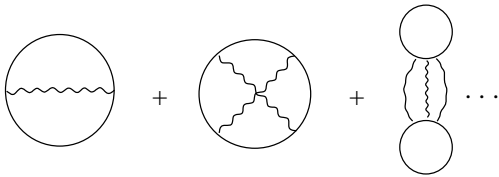}
\caption{Diagrammatic expansion for the effective potential}
\end{center}
\end{figure}

$V_2(S,D)$ is the sum of all two-particle
irreducible vacuum diagrams, depicted in Fig.1, and
the equation
\be
\frac{\d V}{\d S}=0,
\l{delv}
\ee
gives the SDE for fermions. We are not considering the contributions
to the vacuum due to gauge and ghosts loops, because we are interested
only in the vacuum value of the fermionic operator.

We can represent $V_2(S,D)$ analytically in the Hartree-Fock approximation by
\be
\imath V_2(S,D) = - \frac{1}{2} Tr(\G S \G S D)
\l{v2full}
\ee
where $\G$ is the fermion proper vertex. In Eq.(\ref{v2full}) we
have not written the gauge and Lorentz indices, as well as the
momentum integrals.

We want to determine the vacuum expectation value for technifermion self-energy.
Therefore it is better to compute the vacuum energy density, which is given by
the effective potential calculated at minimum subtracted by its
perturbative part, which does not contribute to dynamical
mass generation\cite{cornja,cornor}
\be
\< \Omega \> = V_{min}(S,D) - V_{min}(S_p,D_p),
\l{omega}
\ee
where $S_p$ is the perturbative counterpart of $S$. $V_{min}(S,D)$ is
obtained substituting the SDE Eq.(\ref{delv}) back into Eq.(\ref{vfull}),
and in the chiral limit $S_p = S_0$. The complete fermion propagator
$S$ is related to the free propagator by
\be
S^{-1} = S_0^{-1} - \S ,
\l{sdf}
\ee
where $S_0 = \imath /\not \! p$.

Choosing Landau gauge and working in the Euclidean
space ($P^2 \equiv -p^2$) we find that $V_{min}$ is equal to~\cite{castorina}
\be
V_{min} = 2 N\int \frac{d^4P}{(2\pi)^4} \,
\left[ - \ln ( \frac{P^2 + \S^2}{P^2} ) +
\frac{\S^2}{P^2 + \S^2} \right],
\l{vminf}
\ee
where N is the number of technicolors (techniquarks are in the fundamental representation of SU(N)).
Since we are interested in the vacuum value $\< \Omega \>$, and, particularly,
in its leading term, we can expand $V_{min}$ in powers of $\S/P$ to obtain
\be
\< \Omega \> \simeq -N \int \frac{d^4P}{(2\pi)^4} \frac{\S^4}{P^4}.
\l{omega2}
\ee
To perform the integral in Eq.(\ref{omega2}),
with the self energy given by Eq.(\ref{sa}),
it is helpful to use a particular Mellin transform\cite{cs}
\be
\left[ 1 + \kappa \ln \frac{p^2}{\mu^2} \right]^{-\epsilon} =
\frac{1}{\Gamma ({\epsilon})}\int_0^\infty d\sigma \, e^{-\sigma}
\left( \frac{p^2}{\mu^2} \right)^{-\sigma \kappa} \sigma^{\epsilon - 1}.
\label{mt}
\ee

Before we calculate the expression of the vacuum energy for Eq.(\ref{sa})
it is instructive to show the result for Eq.(\ref{si}) and (\ref{sr}), which
are
\be
\< \Omega \>_I \simeq -\frac{\m^{4}d}{64\pi^{2}b}{\frac{N}{g^{2}}}[1 + bg^{2}(\m^{2})ln(\L^2/\m^2)]^{-4\g},
\label{omega2I}
\ee
\be
\< \Omega \>_R \simeq -\frac{\m^{4}N}{64\pi^{2}}\left(\frac{\m^8}{\L^8}\right)[1 + bg^{2}(\m^{2})ln(\L^2/\m^2)]^{4\g}
\label{omega2R}
\ee
\noindent where $d = \frac{1}{(\gamma - 1/4)}$, $g^{2}=g^{2}(\Lambda^{2})$ and $\Lambda$ is an infrared cutoff momentum.
The natural value of this infrared cutoff is $\Lambda = \mu$ \cite{castorina}. Therefore
\be
\< \Omega \>_I \simeq -\frac{\m^{4}d}{64\pi^{2}b}{\frac{N}{g^{2}}},
\label{omega2Ib}
\ee
\be
\< \Omega \>_R \simeq -\frac{\m^{4}N}{64\pi^{2}}.
\label{omega2Rb}
\ee
It is obvious that the values of the vacuum energy for the ansatz of
Eq.(\ref{sa}) will also interpolate between the ones of Eq.(\ref{omega2Ib}) and
(\ref{omega2Rb}).

There are some interesting remarks about this result. Only a solution behaving
logarithmically as $\S_I$ will give a vacuum
energy proportional to $1/g^2$, any other solution falling off faster
(as $1/p$ at some power) will produce a vacuum energy independent of
g. This is a consequence of the almost convergent behavior of Eq.(\ref{omega2}).
It is possible to verify that the vacuum energy (\ref{omega2}) is related to the fermion condensate
($\< \Omega \> \propto \langle \bar{\psi} \psi\rangle$) (see, for instance, Ref.\cite{nago}),
consequently this one will also be proportional to $1/g^2$.
Therefore for this solution an argument by Gupta and Quinn\cite{quinn} could imply that
this solution cannot be predicted by OPE, because the vacuum value of an operator inversely
proportional to the coupling constant destroys the perturbative power counting of
the OPE series. Finally, the solution $\S_I$ minimizes the vacuum energy unless
the coupling constant in the infrared goes to a very large value, but there are
several indications that this does not happen in QCD\cite{aguilar} and this is also
what could be expected in walking TC theories. Here we shall
restrict ourselves to the scaling law $\frac{g^2}{4\pi} \approx 1$.  We can now
proceed to the calculation of the vacuum energy using Eq.(\ref{sa}), and we
can foresee two possible behaviors: a) the ansatz lead to some intricate
behavior with one or more minima between $\< \Omega \>_I$ and $\< \Omega \>_R$,
b) the vacuum energy interpolates smoothly between $\< \Omega \>_I$ and
$\< \Omega \>_R$ and the first one is the only minimum.

Due to the form of Eq.(\ref{sa}) it is better to write Eq.(\ref{omega2}) in the following
form
\be
\<\Omega\> \simeq -\frac{\m^{4}N}{16\pi^2}\int^{\infty}_{\m^2} \frac{dP^{2}}{P^2}{\S(P^{2})}^{4},
\ee
\noindent where $\S(P^{2})=\S(P^{2})_{TC}/\m$ and using Eq.(\ref{sa}) and the Mellin transform Eq.(\ref{mt}) we obtain
\be
\<\Omega\> \simeq -\frac{\m^{4}N}{16\pi^2}\frac{1}{\Gamma[\d]}\int_{0}^{\infty}{dz}z^{\d - 1}
e^{-z}\int_{1}^{\infty}{dx}x^{-az-4\a -1},
\ee
\noindent where we defined $\d=4\beta{cos}(\a\pi)$ ,  $\beta= 3c/16\pi^{2}b$,  $x = \frac{P^2}{\m^2}$,
and $a = b g^2$. Performing the $x$ integration we obtain
\be
\<\Omega\> \simeq -\frac{\m^{4}N}{16\pi^2}\frac{1}{\Gamma[\d]}\int_{0}^{\infty}{dz}z^{\d - 1}e^{-z}\frac{1}{4\a + az}
\label{Ip}
\ee

We will present our analysis of $\<\Omega\>$ in the different regions of the parameter $\alpha$.
We start with the case $\a \simeq 0$ where we can make the expansion
$$
\frac{1}{4\a + az}\simeq \frac{1}{az}\left[ 1 - \frac{4\a}{az} + ...\right],
$$
\noindent than Eq.(\ref{Ip}) can be put in the following form
\be
\<\Omega\>_0 \simeq -\frac{\m^{4}N}{16\pi^2a}\frac{1}{\Gamma[\d]}\int_{0}^{\infty}{dz}z^{\d - 1}e^{-z}
\frac{1}{z}\left[ 1 - \frac{4\a}{az} + ...\right].
\ee
\noindent Retaining only the first two terms after integration and using properties of Gamma functions we
can write
\be
\<\Omega\>_0 \simeq -\frac{\m^{4}N}{16\pi^2a}\left[\frac{1}{(\d -1)} - \frac{4\a}{a}\frac{1}{(\d -1)(\d - 2)} + ...\right].
\ee
\noindent In this case we have $\d \simeq 4\beta$. The coefficient  $a=bg^2$ can be roughly estimated around of one,
$bg^{2} \approx 1$, because for many groups the coefficient $b$ is always smaller that one. For example,
in the case of $SU(N)$ and $N=4$, we have $b=0.067$ and if we assume the scaling law $\frac{g^2}{4\pi}\approx 1$,
we obtain $bg^2\approx 0.85$\,. For $N=8$ we have $bg^2\approx 2.0$. Therefore we can define
\br
\Phi(\Omega) \equiv \frac{64\pi^2{\<\Omega\>_{0}}}{\m^4N}=&&-\frac{1}{(\beta -1/4)}\times \nonumber\\
&&\left[1 - \frac{4\a}{(\beta - 1/2)} + ...\right].
\er
\noindent
\par This result is plotted in Fig.(2) as a function of $\a$ and $\beta$. In this figure we see that the deeper minimum happens for the smallest value of $\a$ and $\beta$$(=1/2)$;
\begin{figure}[ht]
\begin{center}
\hspace{-2cm}
\epsfig{file=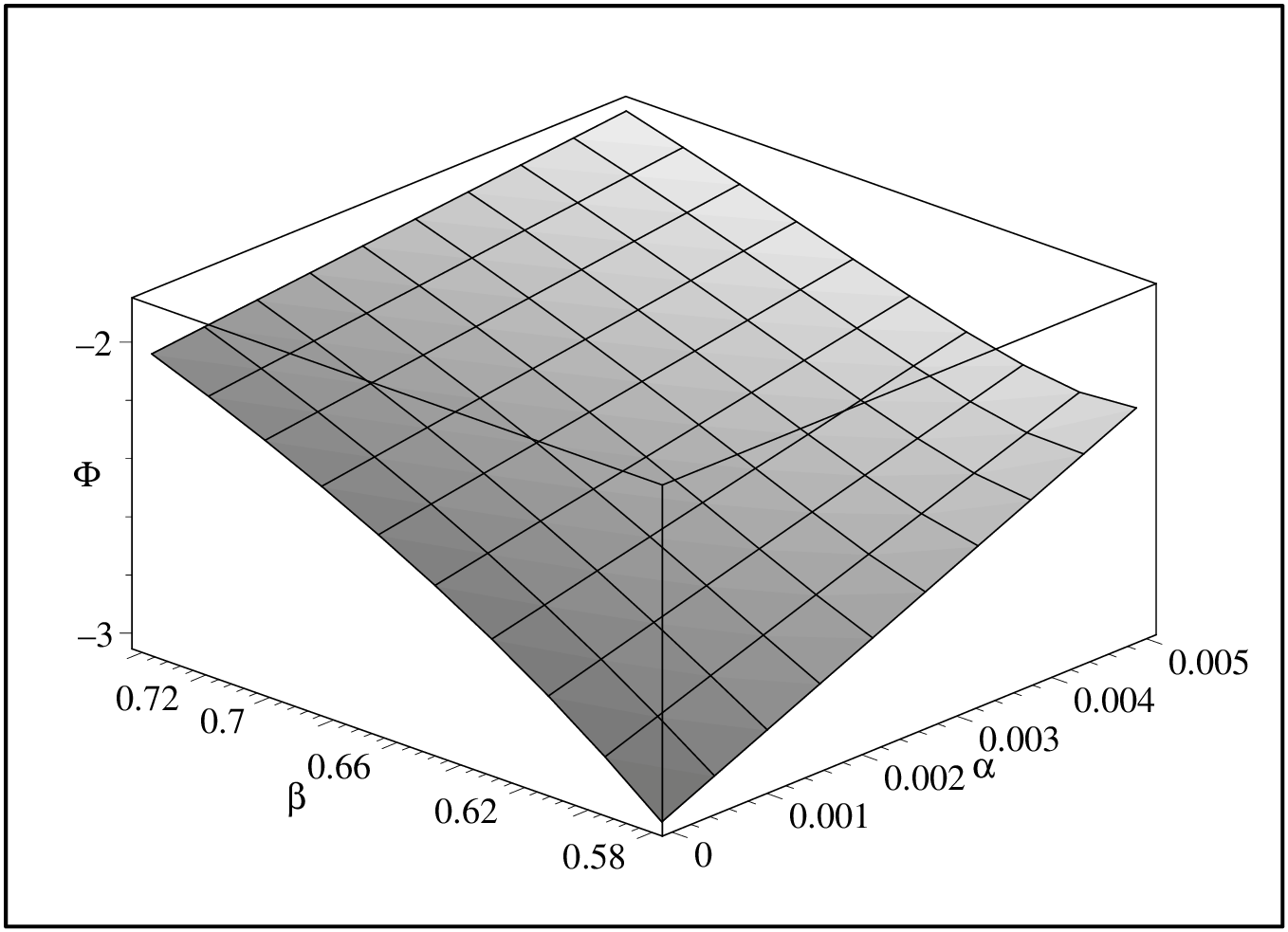,width=0.45\textwidth,angle=-90}
\vspace{-1.5cm}
\caption{Behavior of $ \Phi (\Omega) \equiv \frac{64 \pi^2  \< \Omega \>_{0} }
{ \m^4N} $ plotted in terms of the
$\alpha$ and $\beta$ parameters.}
\end{center}
\end{figure}

\par In the case when $\a \simeq 1$ whe have
\be
\<\Omega\>_1 \simeq -\frac{\m^{4}N}{64\pi^2\a}\left[1 + \frac{a\beta}{\a} + ...\right].
\ee
\noindent Now $\beta cos(\a\pi)\approx -\beta$, and we define
\be
\Gamma(\Omega)\equiv\frac{64\pi^2{\<\Omega\>_{1}}}{\m^4N}=-\frac{1}{\a}\times\left[1 + \frac{\beta}{\a} + ...\right].
\ee
\noindent This case is depicted in Fig.(3).
\begin{figure}[ht]
\begin{center}
\hspace{-2cm}
\epsfig{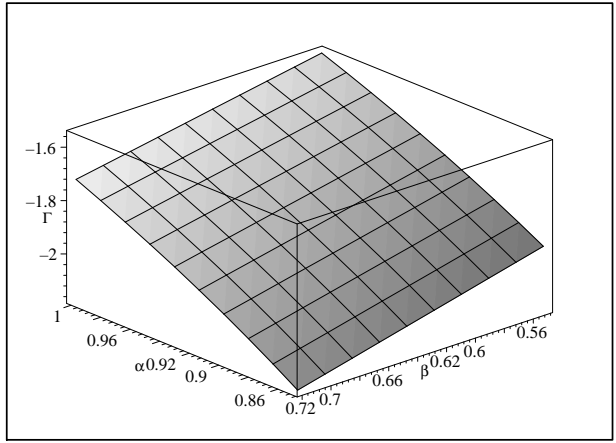}
\vspace{-1.5cm}
\caption{Behavior of $\Gamma(\Omega)\equiv\frac{64\pi^2{\<\Omega\>_{1}}}{\m^4N} $
plotted in terms of the
$\alpha$ and $\beta$ parameters.}
\end{center}
\end{figure}

\noindent Note that in Fig.(3) the region of large $\a$ is one of maximum energy.
The surface decreases smoothly towards small $\a$ changing gradually its
minima from large $\beta$ to small $\beta$;

\par Note that Eq.(\ref{Ip}) can also be solved without recurring to expansions
in terms of the Wittaker functions $(W_{a,b}(x))$ and the result is equal to
\be
\<\Omega\>_F \simeq -\frac{\m^{4}}{16\pi^2a}\left[{\(\frac{4\a}{a}\)}^{\frac{\d - 2}{2}}e^{2\a/a}
W_{-\frac{\d}{2},\frac{1-\d}{2}}(4\a/a)\right].
\ee
\noindent However this solution it is valid only for positive $\beta cos(\a\pi)$. Expanding this solution for $\a \approx 0$ we recover the result for $\<\Omega\>_{0}$.

Our result shows that the deepest minimum of energy happens for a solution
behaving like
\be
\S_{TC}(-p^{2}) \simeq \mu[1 + bg^2(\mu^2)ln(-p^2/\mu^2)]^{-\g} ,
\label{tci}
\ee
where $\g = 3c/16\pi^{2}b$ has the smallest possible value ($\simeq 1/2$).

To conclude we can remember that the expression of Eq.(\ref{tci}) is indeed the
one that solve many of the TC problems. Unfortunately we verified only one
criterion that the solution must obey, from this one to the construction of
realistic models (and test of the criterion) there is still a lot of work to do.

\section*{Acknowledgments}

This research was supported by the Conselho Nacional de Desenvolvimento
Cient\'{\i}fico e Tecnol\'ogico (CNPq) (AAN), by Fundac\~ao de Amparo \`a
Pesquisa do Estado de S\~ao Paulo (FAPESP) (AD,AAN), and by Programa de
Apoio a N\'ucleos de Excel\^encia (PRONEX).

\end{document}